\documentclass[letter]{aa} 

\usepackage{graphicx}
\usepackage{txfonts}
\def\mgii{Mg\,{\sc ii}}
\def\oiii{[O\,{\sc iii}]}

\def\civ{C\,{\sc iv}}
\def\ciii{C\,{\sc iii}}
\def\siiv{Si\,{\sc iv}}
\def\aliii{Al\,{\sc iii}}

\def\feii{Fe\,{\sc ii}}

\usepackage{csquotes}
\usepackage{lineno}
\usepackage{float}

\usepackage{array}      
\usepackage{lipsum}   
\begin{document}

   \title{BAL Outflows are common in Low Eddington Ratio AGN}


   \author{M. Vivek
          \inst{1}\inst{2}
          \and
          Dominika Wylezalek\inst{2}}

   \institute{Indian Institute of Astrophysics,
Koramangala II block,
Bangalore - 560034, Karnataka, India\\
              \email{vivek.m@iiap.res.in} \and
          Astronomisches Rechen-Institut, Zentrum fur Astronomie der Universitat Heidelberg, Monchhofstr. 12-14, D-69120 Heidelberg, Germany\\
             \email{dominika.wylezalek@uni-heidelberg.de }}

   \date{Received September 15, 1996; accepted March 16, 1997}

 
  \abstract
   {Broad absorption line (BAL) quasars exhibit significant outflows, offering insights into active galactic nuclei (AGN) feedback. While typically associated with high Eddington ratios, BAL quasars also occur in low Eddington ratio regimes, which remain poorly understood. This study aims to compare BAL properties and variability across these regimes.}
   {We investigate the  occurance rates, absorption characteristics, and variability of BAL quasars at low and high Eddington ratios.}
   {Using the SDSS DR16 quasar catalog, we selected a redshift-matched control sample to compare low and high Eddington ratio BAL quasar sources. We first examined the BAL fraction as a function of Eddington ratio. Key absorption parameters—equivalent width, absorption line width, velocity range, and depth—were analyzed, and a multi-epoch variability study was conducted using repeat spectra, followed by a comparison of parameter distributions between the two samples.}
   {For the first time, we report an increase in BAL fraction toward low Eddington ratios, in addition to the previously known trend of high BAL fraction at high Eddington ratios. While high Eddington sources show extreme absorption features, overall distributions are statistically similar except for maximum outflow velocity. No significant variability differences were observed. The correlation between outflow velocity, Eddington ratio, and luminosity supports the role of radiation pressure in driving quasar outflows. For low Eddington ratios, additional mechanisms, such as softer SEDs, larger outflow distances, and thickened accretion disks from radiatively inefficient processes, likely drive outflow formation.}
   {}


   \maketitle
%

\section{Introduction} \label{sec:intro}
Broad absorption line (BAL) quasars, characterized by strong, blue-shifted absorption features, provide a unique window into the powerful outflows emanating from the central regions of quasars. These outflows, thought to be driven by intense radiation from the accretion disk \citep{proga2000}, can significantly impact the surrounding galaxy, potentially regulating star formation and galaxy evolution \citep[see,][for a review]{Fabian2012}. 

The Eddington ratio is a crucial parameter in understanding quasar accretion physics \citep{shen2014}. The Eddington luminosity is the limiting luminosity for perfectly spherical accretion, and the Eddington ratio - the ratio of a quasar’s bolometric luminosity to its Eddington luminosity - effectively measures how close a quasar’s radiation pressure is to counteracting gravitational forces that confine material in the accretion disk.
The Eddington ratio is also linked to Eigenvector 1 (EV1), a prominent principal component in quasar spectral properties, which captures correlations between various emission line strengths, particularly the ratio of \feii\ to \oiii\ emission, along with the prominence of broad H$\beta$ emission \citep{boroson1992}. Eigenvector 1 is often associated with differences in accretion physics and orientation effects, and BAL quasars are thought to represent quasars with strong radiative outflows influenced by near-Eddington accretion. In line with this, \citet{Boroson2002} found that BAL quasars are predominantly located in the high Eddington ratio, high accretion rate corner of the PCA1-PCA2 eigenvector coefficient space.

\citet{Ganguly2007} analyzed a sample of quasars from the Sloan Digital Sky Survey (SDSS) Data Release 2 (DR2) and found that BALs are more prevalent in quasars accreting near the Eddington limit, though they also occur in quasars with only a few percent of Eddington accretion. Additionally, they observed that maximum outflow velocities increase with both luminosity and spectral blueness, supporting a model of outflow acceleration driven by ultraviolet line scattering. Recently, \citet{Leighly2022} reported a bimodal distribution of Eddington ratios in their analysis of a sample of 30 iron low ionization BALs (FeLoBALs), characterized by either high or low values, with intermediate ratios notably absent.   

In this study, we utilize a sample of BAL quasars from the SDSS Data Release 16 \citep[DR16,][]{SDSSDR162020} to investigate whether the properties and prevalence of BAL outflows vary across different Eddington ratios. By dividing our sample into low and high Eddington ratio sources and constructing a control sample matched in redshift, we examine the fraction of BAL quasars as a function of Eddington ratio and perform a comparative analysis of key outflow parameters. Additionally, multi-epoch spectra are analyzed to explore BAL variability and its correlation with Eddington ratio.
\section{Sample Selection}\label{sec:sample}
Our study utilizes quasars from the SDSS DR16 quasar catalog \citep{Lyke2020}, the largest compilation of quasars from the Sloan Digital Sky Survey, containing over 750,000 spectroscopically confirmed quasars. To investigate the properties of BAL quasars across different Eddington ratios, we cross-matched the SDSS DR16 quasar catalog with the catalog by \cite{Wu2022}, which provides a comprehensive set of derived quasar parameters. The \citet{Wu2022} catalog is particularly valuable for our analysis as it includes parameters obtained from detailed spectral fitting using the PyQSOFit software, allowing us to access accurately derived measurements such as bolometric luminosity, black hole mass, and the Eddington ratio.
\begin{figure}[h]
    \centering
    \includegraphics[width=1\linewidth]{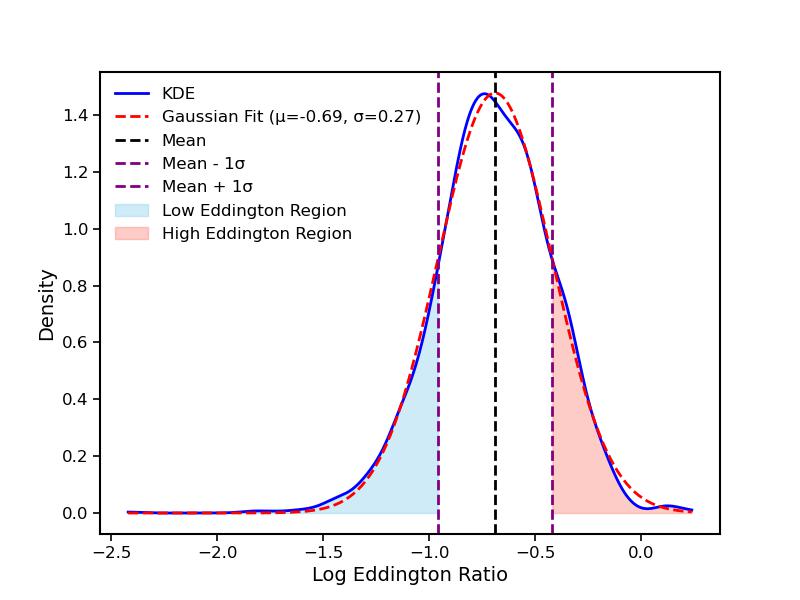}
    \caption{ Distribution of Log Eddington Ratio for the quasar sample, showing the Kernel Density Estimation (KDE) curve (blue solid line) with an overlaid Gaussian fit (red dashed line). The vertical lines indicate the mean (black dashed line) and one sigma deviation from the mean (purple dashed lines). The shaded sky-blue region indicates the low Eddington ratio range, while the shaded salmon region highlights the high Eddington ratio range, defining the boundaries used for sample selection. }
    \label{fig:sample}
\end{figure}
We applied filters to select quasars with reliable Eddington ratio measurements (i.e. sources with black hole mass measurements using \mgii\ line), high signal-to-noise ratios ($>$ 10), and a redshift range of 1.88--2.44. The chosen redshift range   ensures that the prominent emission lines \siiv, \civ, \ciii, \aliii, and \mgii\ are captured within the SDSS spectra, which is essential for accurately identifying BAL features.  We refer to this selection as the main sample.
 We only included black hole mass measurements exclusively based on \mgii\ rather than \civ, as \civ-derived masses are prone to biases and inaccuracies due to line shifts and asymmetries associated with outflows \citep{brotherton2015,coatman2017}. Accurate redshifts in BAL quasars are challenging due to absorbed emission lines, affecting bolometric luminosity estimates. To ensure reliability, we applied a strict cut, requiring visually inspected redshifts \citep[See,][for a discussion on the difference between pipeline and actual redshifts]{rankine2020}.

To categorize quasars into low and high Eddington ratio groups, we fitted a Gaussian function to the kernel density estimate (KDE) distribution of the Log Eddington ratio values for the entire dataset.  The KDE was generated using a Gaussian kernel with a bandwidth of 0.2, determined following the recommended Scott's rule. The Gaussian fit resulted in a mean  of -0.69 with a standard deviation of 0.27. Based on this distribution, we defined low Eddington ratio sources as those with values more than one standard deviation ($\sigma$) below the mean, and high Eddington ratio sources as those exceeding one standard deviation above the mean. This classification resulted in a sample of 326 low Eddington ratio sources and 324 high Eddington ratio sources. Fig.~\ref{fig:sample} shows the distribution of  Eddington Ratio for the quasar sample selected in this work. To identify BAL quasars within the low and high Eddington ratio samples, we used the BAL\_PROB parameter from the SDSS DR16 quasar catalog, selecting sources with a BAL probability greater than 0.5. This criterion yielded 162 BAL quasars in the low Eddington ratio sample and 121 BAL quasars in the high Eddington ratio sample. To create a balanced comparison between the low and high Eddington ratio BAL quasar samples, we constructed a control sample from the low Eddington ratio sources with redshifts closely matched to those in the high Eddington ratio sample.  This approach accounts for any quasar luminosity evolution across redshifts between the two samples. This redshift-matching process resulted in equal sample sizes, with 121 sources in both the low and high Eddington ratio groups.

To investigate BAL variability across different Eddington ratio regimes, we searched for the availability of repeat spectra in each sample. 
Overall, this results in 74 sources with repeat spectra in the low Eddington ratio sample and 102 sources with repeat spectra in the high Eddington ratio sample, providing a  basis for examining BAL variability across multiple observations.

\section{Analysis}\label{sec:analysis}
\subsection{BAL fraction}\label{subsec:bal_frac}
\begin{figure}[h]
    \centering
    \includegraphics[width=1\linewidth]{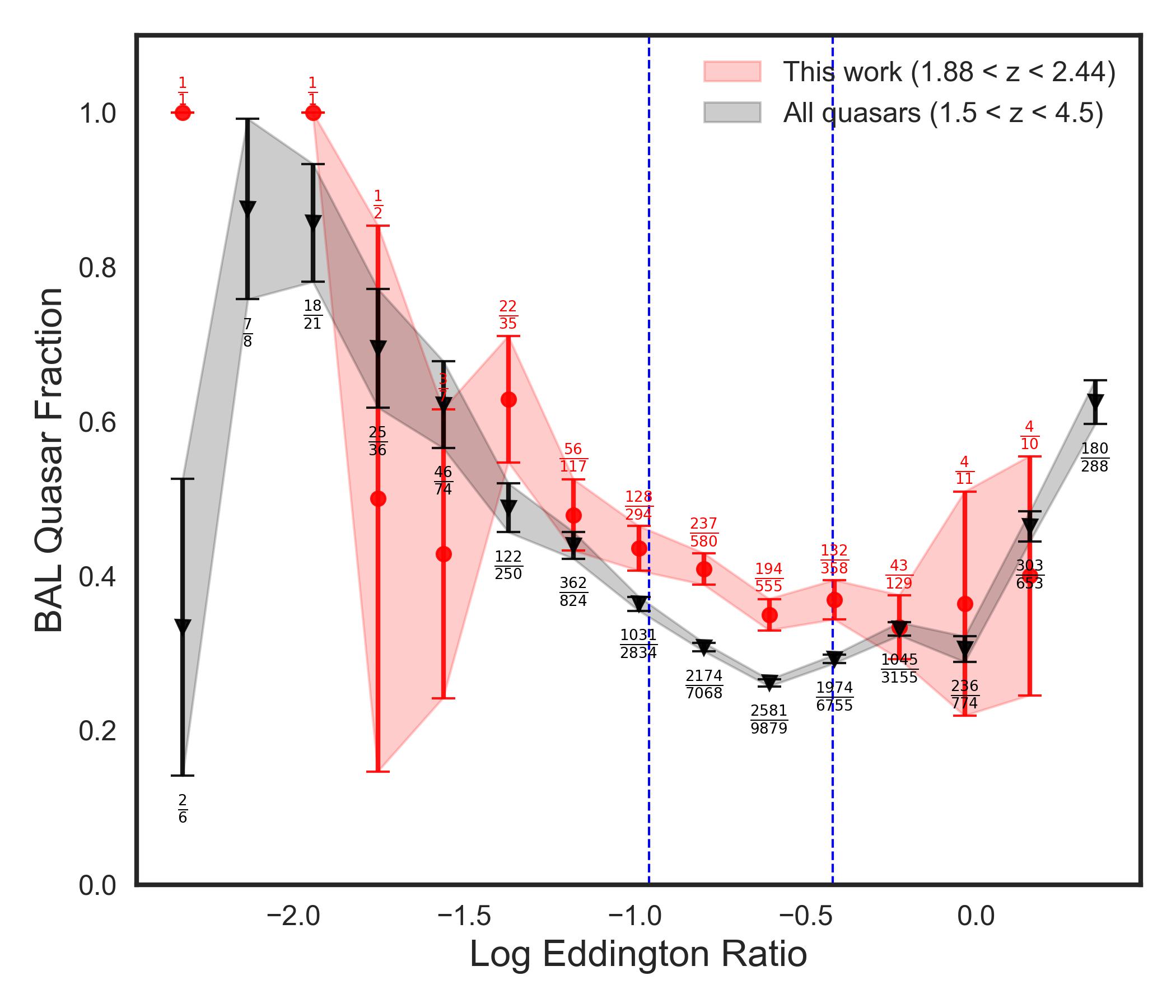}
    \caption{The figure shows the BAL quasar fraction as a function of Log Eddington ratio. Error bars and shaded regions represent binomial errors. Red circles correspond to the 1.88 $<$ z $<$ 2.44 sample with \mgii-based black hole masses, while black triangles represent the broader 1.5 $<$ z $<$ 4.5 sample. Blue dashed lines mark one standard deviation from the mean Log Eddington ratio, defining the low and high Eddington ratio regions.}
    \label{fig:bal_fraction}
\end{figure}

Studying the BAL fraction across different Eddington ratio regimes is essential to understanding the connection between accretion dynamics and the occurrence of quasar outflows.
Fig.~\ref{fig:bal_fraction} shows the fraction of BAL quasars as a function of Log Eddington ratio, with error bars representing the standard binomial error in each bin.  We considered two samples here: the first being the main sample discussed in  section~\ref{sec:sample} (i.e., 1.88 $<$ z $<$ 2.44). Interestingly, the BAL fraction shows an increase at both low and high Eddington ratios in this sample. To further investigate, we examined a broader sample (1.5 $<$ z $<$ 4.5) using \civ\ and \mgii-based black hole mass estimates, with the lower redshift limit ensuring \civ\ remains within SDSS spectral coverage for reliable BAL classification.

The overall BAL fraction is 22\% and 24\% in the 1.5$<$z$<$4.5 and 1.88$<$z$<$2.44 samples, respectively, without any additional filters.   Enforcing a criterion of signal-to-noise$>$10, increases the BAL fraction to 26\% for the 1.5$<$z$<$4.5 sample and 39\% for the 1.88$<$z$<$2.44 sample. Notably, the BAL fraction increases for both low and high Eddington ratio sources, with the trend being more pronounced in the 1.5$<$z$<$4.5 sample due to its larger number of sources, which minimizes the impact of low-number statistics. Regardless of the filters applied, we have confirmed that the trend of increasing BAL fraction toward low Eddington ratio sources persists across the sample (See Appendix~\ref{app_1}, \ref{app_2}). Appendix~\ref{app_23}
presents an analysis of BAL fraction in bins of bolometric luminosity and black hole mass.  Here again, we  observe that BAL fractions tend to increase on either side of Eddington ratios between -0.5 and -1, both at constant luminosity and constant black hole mass. In conclusion, our findings confirm the established increase in BAL fraction at high Eddington ratios while also revealing a new trend: an increase in BAL fraction at low Eddington ratios.

\subsection{BAL parameters}\label{subsec:BALparam}
 We normalized each spectrum in our low and high Eddington ratio samples using either a Principal Component Analysis (PCA) reconstruction or a composite quasar template from \citet{Vandenberk2001} (See Appendix \ref{app_3} for detailed information on the continuum normalization procedures and  the identification of BAL features).

In our analysis, we noticed that sometimes the absorption lines are either too narrow  or too shallow.  However, such features may not meet the conventional definition of broad absorption lines or may arise due to poor continuum fits. To address this, we visually inspected each spectrum to identify genuine BAL cases—those with significantly wide and deep absorption features. We refer to these carefully selected sources as the Pristine sample, while the complete set is termed the Full sample. For low Eddington ratio sources, we identified 88 Pristine BAL quasars, while the high Eddington ratio sources include 84 Pristine BAL quasars.
\begin{figure*}
    \centering
    \includegraphics[scale=0.5]{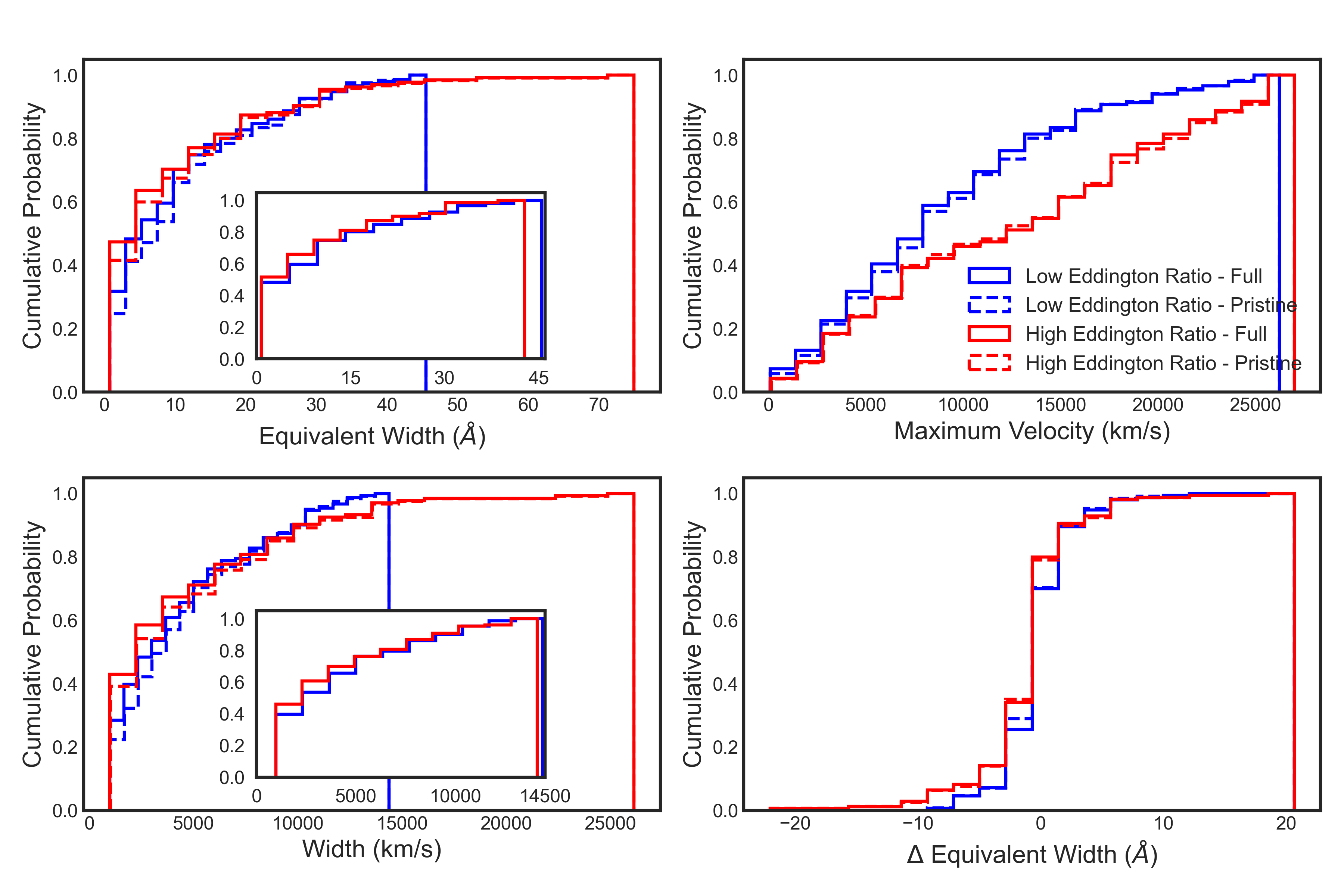}
    \caption{Cumulative probability distributions of equivalent width (upper left), maximum velocity (upper right), absorption line width (lower left), and difference in equivalent width (lower right) for the low (red) and high Eddington ratio (blue) samples. Solid lines represent the full sample, while dashed lines correspond to the pristine sample.  The inset panels show the distributions of low and high Eddington ratio samples after excluding these high-value sources from the high Eddington ratio sample.}
    \label{fig:abs_param}
\end{figure*}

We then measured key absorption parameters for each BAL feature, including equivalent width, absorption line width, minimum and maximum velocity, maximum depth, and mean depth of the absorption lines. These parameters provide a comprehensive characterization of the BAL profiles, capturing both the strength and extent of the absorption features in our sample.

We also measured absorption parameters for multi-epoch spectra as previously described. For continuum normalization of these repeat spectra, we did not fit a new continuum. Instead, we matched the continuum flux of each secondary spectrum to that of the primary spectrum in the original sample, and used the corresponding continuum used for the normalization of the primary spectra.   
We   fixed the same velocity edges when measuring BAL parameters in the repeat spectra to ensure consistent comparisons.  We further computed the differences between the rest-frame timescales, equivalent width, depth, and depth-weighted velocity for each BAL feature across the multi-epoch spectra.

\section{Results}\label{sec:results}
We compared the distributions of the derived parameters—equivalent width, maximum and minimum velocity, absorption line width, mean depth, and maximum depth—between the low and high Eddington ratio samples, for both the full sample and the pristine sample. For the variability analysis, we compared the distributions of the differences in parameters across multi-epoch observations.

The Kolmogorov-Smirnov (KS) and Anderson 2-sample tests yield a p-value below 0.05 only for the minimum and maximum velocities, indicating a statistically significant difference in these parameters between the low and high Eddington ratio samples. However, we observe that the high Eddington ratio sample contains more sources with higher equivalent widths and broader velocity widths, though this difference is not statistically significant. For instance, only three sources in the high Eddington ratio sample have an equivalent width exceeding the maximum equivalent width observed in the low Eddington ratio sample.

{For the difference distributions in the variability data, we observed that the low Eddington ratio sample had slightly more points at small MJD differences (i.e., probed rest-frame timescales) as compared to the high Eddington ratio sample.} The KS and Anderson tests yielded a p-value below 0.05 for the MJD-difference parameter, suggesting that the probed timescales differ between the samples. This difference in probed timescales implies that parameter differences, like equivalent width difference cannot be directly compared between the low Eddington ratio sample and high Eddington ratio sample. To address this, we further controlled the MJD-difference in the low Eddington ratio sources to match that of the high Eddington ratio sources. After this, the KS and Anderson tests showed no statistically significant differences in the distributions of the variability parameters for low and high Eddington ratio sources. However, as with the equivalent width distributions, we note that the extreme cases of equivalent width differences are primarily in the high Eddington ratio sample.

Fig.~\ref{fig:abs_param}  shows the cumulative probability distributions of equivalent width (upper left), maximum velocity (upper right), absorption line width (lower left), and difference in equivalent width (lower right) for the low (red) and high Eddington ratio (blue) samples. Solid lines denote the full sample, while dashed lines represent the pristine sample. There is no significant difference between the full and pristine samples for any parameter. As discussed, the high Eddington ratio sample shows a few sources with particularly large values for equivalent width and absorption line width. The insets in these panels compare the distributions of low and high Eddington ratio samples after excluding these high-value sources from the high Eddington ratio sample.  Consistent with the KS and Anderson tests, Fig.~\ref{fig:abs_param} shows no notable differences in any parameters except for maximum velocity, which remains distinct between the two Eddington ratio groups.

We conclude that the low Eddington ratio sample exhibits a similar distribution of absorption line properties and absorption line variability to the high Eddington ratio sample, with the exception of the maximum velocity parameter. Previous studies, such as \citet{Ganguly2007}, have also noted a connection between Eddington ratio and maximum outflow velocity. Our comparative analysis between low and high Eddington ratio sources confirms this trend, further supporting the relationship between Eddington ratio and maximum outflow velocity.

\section{Discussion \& Conclusion}\label{sec:discussion}
\begin{figure}
    \centering
    \includegraphics[width=0.85\linewidth]{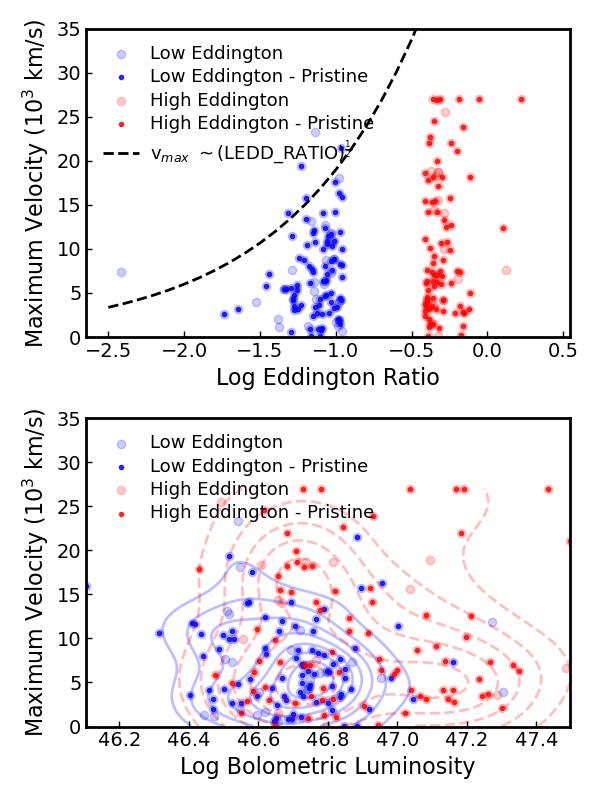}
    \caption{Top Panel : Correlation between the maximum velocity of absorption and the
Eddington ratio. The black dashed line represents a fiducial curve to show the predicted scaling relationship from Hamann(1998) and Misawa et al. (2007), with arbitrary normalization. Bottom Panel : Correlation between velocity of absorption and the bolometric luminosity, with overlaid contour plots for enhanced clarity. }
    \label{fig:scatter}
\end{figure}

Our study provides a comparative analysis of BAL quasars across low and high Eddington ratio regimes. Notably, we observe an increase in BAL fraction at both low and high Eddington ratios. While \citet{Ganguly2007} reported that BALs are more common in quasars accreting near the Eddington limit, they also noted their presence in sources accreting at a small fraction of the Eddington rate. Our analysis, however, highlights a distinct increase in BAL fraction toward low Eddington ratios, in the restrictive sample (1.88 $<$ z $<$ 2.44) using \mgii-based black hole masses. This trend is even stronger in the broader sample (1.5 $<$ z $<$ 4.5) incorporating masses derived from \mgii\ or \civ\ emission lines.  We also note that \citet{Ganguly2007}'s sample features more high Eddington ratio sources, whereas our sample includes a larger proportion of low Eddington ratio sources (see Appendix \ref{app_2} for details).

While \civ-based mass estimates are debated due to non-virial motions like outflows that can overestimate masses and distort Eddington ratios, the trend observed in our conservative sample using \mgii-based masses supports the conclusion that BAL quasars are as common, if not more so, in low Eddington ratio sources. Supporting this, \citet{Leighly2022} reported that FeLoBAL quasars exhibit either high or low Eddington ratios, with intermediate values absent.  

In terms of absorption line properties, we find that parameters such as equivalent width, line width, and depth are remarkably similar for both low and high Eddington ratio samples.  This indicates that the basic characteristics of BAL strength do not significantly depend on the accretion rate, although the most extreme cases in our sample—those with the highest equivalent width, and width—are associated with high Eddington ratio sources.

The maximum outflow velocity distribution differs significantly between high and low Eddington ratio sources, with higher Eddington ratio sources exhibiting larger outflow velocities. This result aligns with our composite spectra analysis (see Appendix \ref{app_4}), which also shows higher outflow velocities for high Eddington ratios. According to \citet{hamann1998,misawa2007}, the terminal velocity of quasar outflows scales with the Eddington ratio as \( v_{\text{terminal}} \propto (L_{\text{bol}} / L_{\text{Edd}})^{1/2} \), indicating a link between Eddington ratio and maximum outflow velocity.  This relationship assumes a constant force multiplier and a fixed launch radius for all sources. 
\citet{Ganguly2007} observed a similar scaling of \( v_{\text{max}} \) with Eddington ratio, along with correlations between maximum outflow velocity, the luminosity at 3000 $\AA$, and the spectral index—supporting radiation pressure-driven outflows.    Likewise, \citet{Laor2002} found a similar trend between normalised velocities, v$_{\text{max}}$ / v$_{\text{BLR}}$, and Eddington ratios, but with a steeper slope of 0.83.  Fig.~\ref{fig:scatter} illustrates these relationships, showing a clear correlation between \( v_{\text{max}} \), Eddington ratio, and bolometric luminosity. While the Eddington ratio imposes an upper limit on outflow velocities, the trend is less distinct for bolometric luminosity, although high-luminosity sources tend to exhibit larger velocities, as evident from contour overlays.
These findings tentatively support the role of radiation pressure in driving quasar outflows, consistent with theoretical models predicting higher outflow velocities in sources with high Eddington ratios.

While the general variability characteristics, including changes in equivalent width and depth over time, are similar for both low and high Eddington ratio samples, the most extreme cases of variability are seen with high Eddington ratio sources. \citet{Wilhite2008} demonstrated that quasar continuum variability is anti-correlated with Eddington ratio, indicating that lower Eddington ratio sources generally exhibit greater continuum variability. Given this, our finding that both low and high Eddington ratio sources show similar absorption line variability suggests that the mechanisms driving changes in absorption lines may not always be directly linked to continuum variability. Instead, absorption line variability could be influenced by other factors, such as changes in the outflow structure, covering fraction, or line-of-sight conditions.

In our sample, the luminosity distribution remains comparable, while the black hole masses are an order of magnitude larger for low Eddington ratio sources.  The significantly larger black hole mass likely produces a softer SED \citep{Laor2011}, with an excess of UV photons relative to X-ray photons, potentially mitigating the over-ionization issue in outflows \citep{proga2000}.  However, \citet{Vito2018} reported no correlation between X-ray weakness and Eddington ratio in a sample of 30 BAL quasars.

A possible connection to changes in the accretion disk geometry remains unexplored. At both extremes of Eddington ratios—typically considered to be around 1\% of the Eddington limit on the lower side and close to the Eddington limit (unity) on the higher side—the standard Shakura-Sunyaev thin disk model \citep{Shakura1973} may not fully apply, as the disk geometry is likely altered by radiatively inefficient processes.  In these cases, the inner accretion disk puffs up, adopting a slim disk structure for high Eddington ratio sources or a thick structure associated with hot accretion flows in low Eddington ratio sources \citep{Yuan2014}, both of which reduce vertical support for the gas. This modified geometry allows turbulence to more easily lift gas above the disk plane, where radiation pressure can then accelerate it outward, driving strong outflows. Another favorable factor is the efficient shielding offered by the altered inner accretion disk, which helps prevent overionization in the outflows. 
Consequently, the observed increase in BAL fraction at both low and high Eddington ratios may result from this altered disk structure, which enhances conditions for launching outflows across a wide range of accretion rates.  Additionally, other processes, such as magnetic or thermal driving, may also contribute to launching and sustaining outflows in low Eddington ratio sources.

\begin{acknowledgements}
The authors thank the anonymous referee for the constructive comments that have helped to improve the manuscript.    MV acknowledges support from DST-SERB in the form of a core research grant (CRG/2022/007884). DW acknowledges support through an Emmy Noether Grant of the German Research Foundation. MV and DW acknowledge support through the MERAC Foundation.

\end{acknowledgements}

\bibliography{references}{}
\bibliographystyle{aasjournal}

\appendix
\section{BAL Fraction Across Varying S/N Thresholds and BAL Definitions}\label{app_1}
\begin{figure*}
    \begin{tabular}{ccc}

       \includegraphics[scale=0.38,trim={5pt 0pt 5pt 5pt}, clip]{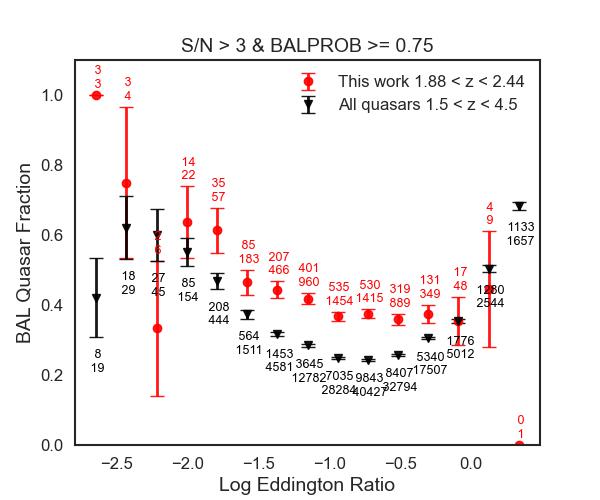}  &
        \includegraphics[scale=0.38,trim={5pt 0pt 5pt 5pt}, clip]{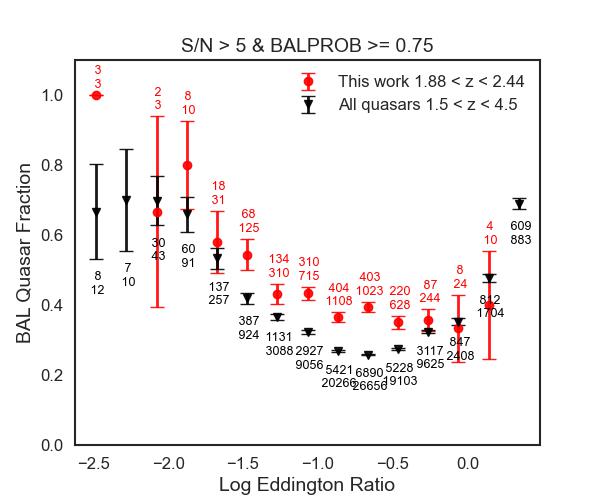}  &
         \includegraphics[scale=0.38,trim={5pt 0pt 5pt 5pt}, clip]{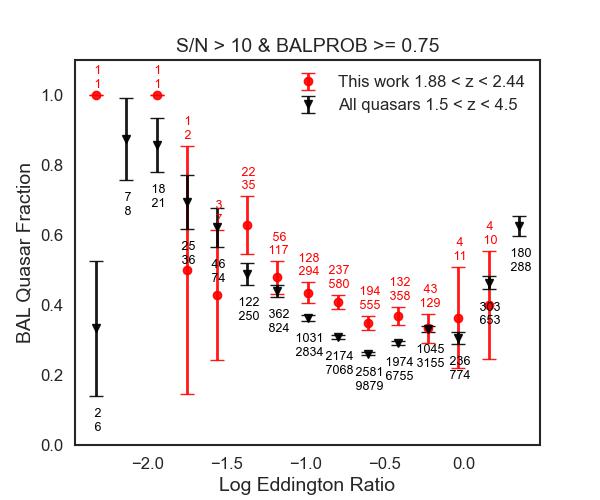}   \\
         
     \includegraphics[scale=0.38,trim={5pt 0pt 5pt 5pt}, clip]{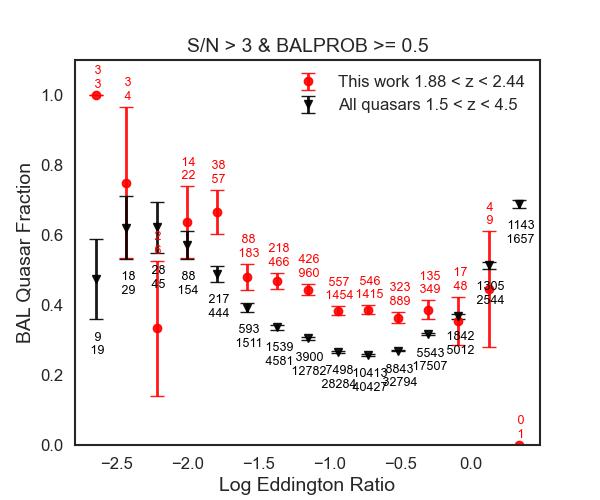}    &
     \includegraphics[scale=0.38,trim={5pt 0pt 5pt 5pt}, clip]{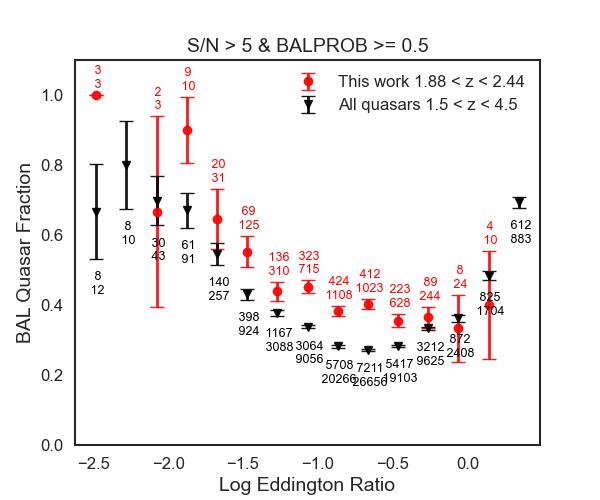}    &
     \includegraphics[scale=0.38,trim={5pt 0pt 5pt 5pt}, clip]{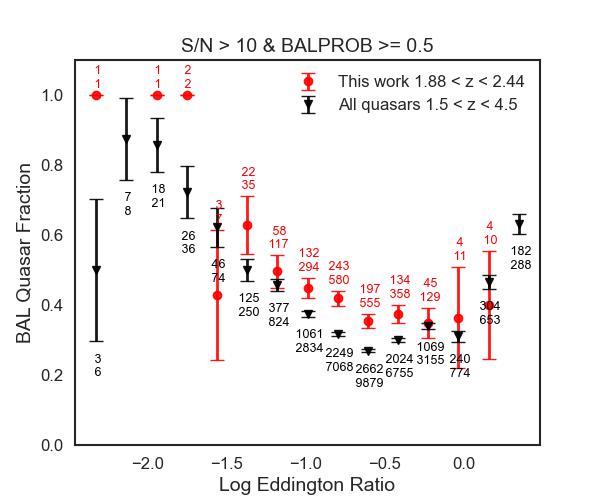}    \\

    \end{tabular}
    \caption{The figure shows the BAL fraction across two quasar samples analyzed in this study: (1) quasars with redshifts between (1.88 $<$ z $<$ 2.44) with reliable Eddington ratio estimates based solely on \(\text{Mg II}\) black hole mass measurements, and (2) quasa), with all available Eddington ratio measurements. The panels are arranged in rows and columns to showcase the effects of different signal-to-noise (S/N) cuts and BAL definitions. Each column represents a distinct S/N threshold (top: S/N $>$ 3, middle: S/N $>$ 5, bottom: S/N $>$ 10), while each row corresponds to a different BAL definition from the DR16 quasar catalog. The definitions vary by BAL probability: \( \text{BAL\_PROB} \geq 0.75\) (includes cases with significant Absorption Index measurements), and \( \text{BAL\_PROB} \geq 0.5\) (includes  all BAL quasars including the ones with less significant Absorption Index measurements).}
    \label{fig:snr_bal_prob}
\end{figure*}

 To assess the impact of data quality and BAL definitions, we analyze these samples under varying signal-to-noise (S/N) thresholds and BAL criteria. Specifically, we use three S/N cuts (S/N $>$ 3, S/N $>$ 5, and S/N $>$ 10) and two levels of BAL probability from the DR16 quasar catalog.  BAL quasars with \( \text{BAL\_PROB} \geq 0.75\) include those with significant Absorption Index values, while  \text{BAL\_PROB} $\geq$ 0.5 encompasses a broader set, allowing for BAL quasars with less significant Absorption Index measurements.  Fig.~\ref{fig:snr_bal_prob} shows the BAL fraction for two samples considered in this study;  (1) quasars within the redshift range (1.88 $<$ z $<$ 2.44) with reliable Eddington ratio estimates derived solely from \mgii-based black hole mass measurements, and (2) a broader sample of quasars with redshifts between (1.5 $<$ z $<$ 4.5), encompassing all available Eddington ratio measurements. Each column represents a distinct S/N threshold (top: S/N $>$ 3, middle: S/N $>$ 5, bottom: S/N $>$ 10), while each row corresponds to a different BAL definition from the DR16 quasar catalog. The number of BAL quasars and the total number of quasars in each bin are annotated on the plot. The results consistently reveal an increase in the BAL fraction within the lower Eddington ratio bins, regardless of the S/N threshold or BAL definitions used. 
\section{comparison  with Ganguly et al 2007} \label{app_2}
\begin{figure*}
    \centering
    \begin{tabular}{cc}

    \includegraphics[width=0.5\linewidth]{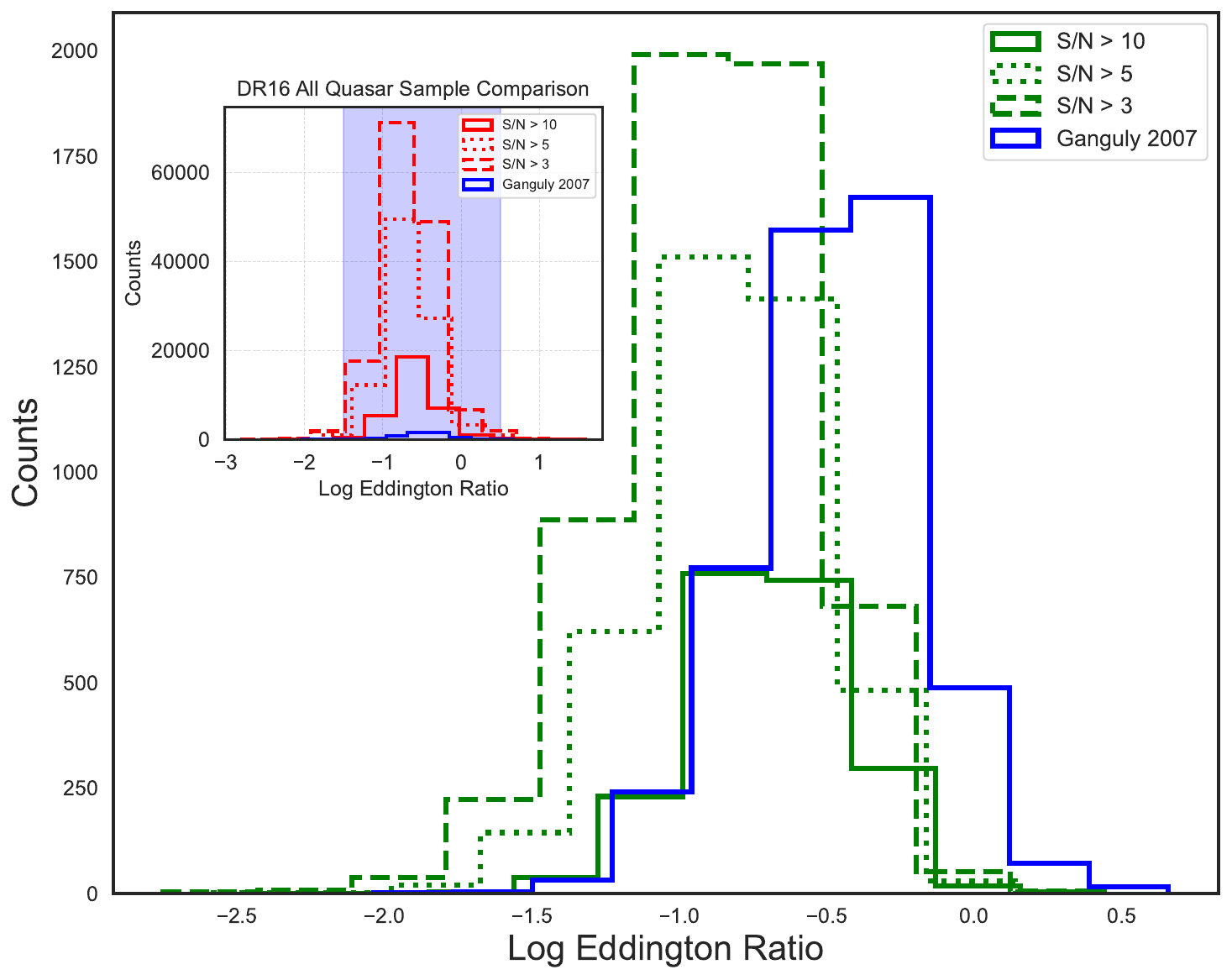}
            & 
            \includegraphics[width=0.5\linewidth]{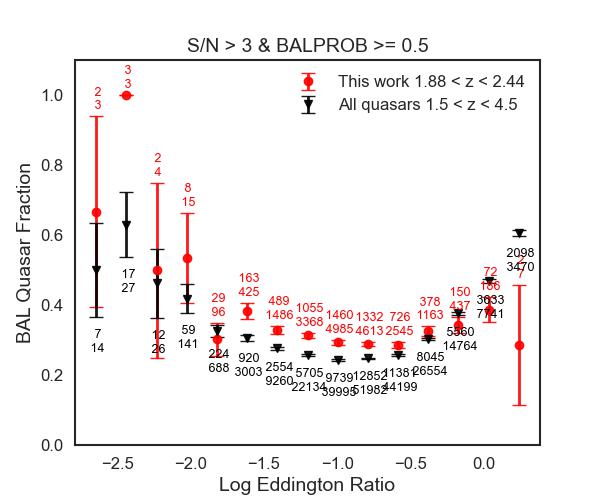}
    \end{tabular}
    \caption{Left Panel :  The distribution of Eddington ratios for the two quasar samples in this study, as compared to \citet{Ganguly2007}. Green curves represent quasars within \(1.88 < z < 2.44\) with reliable \mgii-based Eddington ratios across three S/N thresholds, while the \citet{Ganguly2007} sample appears in blue. The inset shows a broader quasar sample with \(1.5 < z < 4.5\) and all available Eddington ratios, with the shaded region indicating the \citet{Ganguly2007} Eddington ratio distribution. Right Panel : The BAL fraction as a function of log Eddington ratio for the S/N > 3 threshold sample with just \texttt{ZWARNING == 0} selection filter.}
    \label{fig:lambda_distribution}
\end{figure*}

The left panel of fig.~\ref{fig:lambda_distribution} shows the distribution of Eddington ratios for the two quasar samples examined in this study, alongside the distribution from \citet{Ganguly2007}. The green curves represent the histogram counts for quasars within the redshift range (1.88 $<$ z $<$ 2.44) with reliable Eddington ratio estimates derived solely from \mgii-based black hole mass measurements, shown across three different S/N thresholds. In contrast, the \citet{Ganguly2007} sample is shown in blue. The inset panel displays histogram counts for a broader quasar sample with redshifts between (1.5 $<$ z $<$ 4.5), incorporating all available Eddington ratio measurements. In the inset, the shaded region corresponds to the Eddington ratio distribution from \citet{Ganguly2007}.

Our sample extends significantly to lower Eddington ratios, while the \citet{Ganguly2007} analysis includes a larger proportion of high Eddington ratio sources.  Notably, our sample lacks high Eddington ratio sources, which may explain why the increase in BAL fraction is not as pronounced for high Eddington ratios as compared to lower ones. While \citet{Ganguly2007} analyzed a sample of 4,858 quasars within the redshift range \(1.7 < z < 2\), our reliable Eddington ratio sample contains a comparable number of sources across different S/N thresholds: 2,101 quasars for S/N $>$ 10, 4,235 for S/N $>$ 5, and 5,867 for S/N $>$ 3. For the broader sample, which includes all available Eddington ratios, our study includes a substantially larger number of quasars, with 32,830 sources for S/N $>$ 10, 94,762 for S/N $>$ 5, and 148,887 for S/N $>$ 3. This significantly expanded sample size, particularly with an increased representation at lower Eddington ratios, enables a more comprehensive examination of Eddington ratio distributions and BAL fraction trends. The higher number of high Eddington ratio sources in \citet{Ganguly2007} likely arises because they used data from SDSS DR2, which primarily included bright quasars. Subsequent BOSS and eBOSS surveys focused on fainter targets, often excluding these brighter quasars. Additionally, our study imposes a strict condition that quasars must be visually inspected for reliable redshift estimates, thus relying heavily on the DR14 quasar catalog \citep{paris2018}. 
To mitigate this limitation, we relaxed the redshift criteria and included all quasars with \texttt{ZWARNING == 0}. The right panel of fig~\ref{fig:lambda_distribution} shows the BAL fraction as a function of Log Eddington ratio for S/N $>$ 3 threshold with just \texttt{ZWARNIGN == 0} selection filter. This again supports our finding of an increased BAL fraction at both low and high Eddington ratios.

\section{BAL fraction as a function of Black hole mass \& Luminosity}\label{app_23}
\begin{figure*}
\begin{tabular}{cc}
 \includegraphics[width=0.5\linewidth]{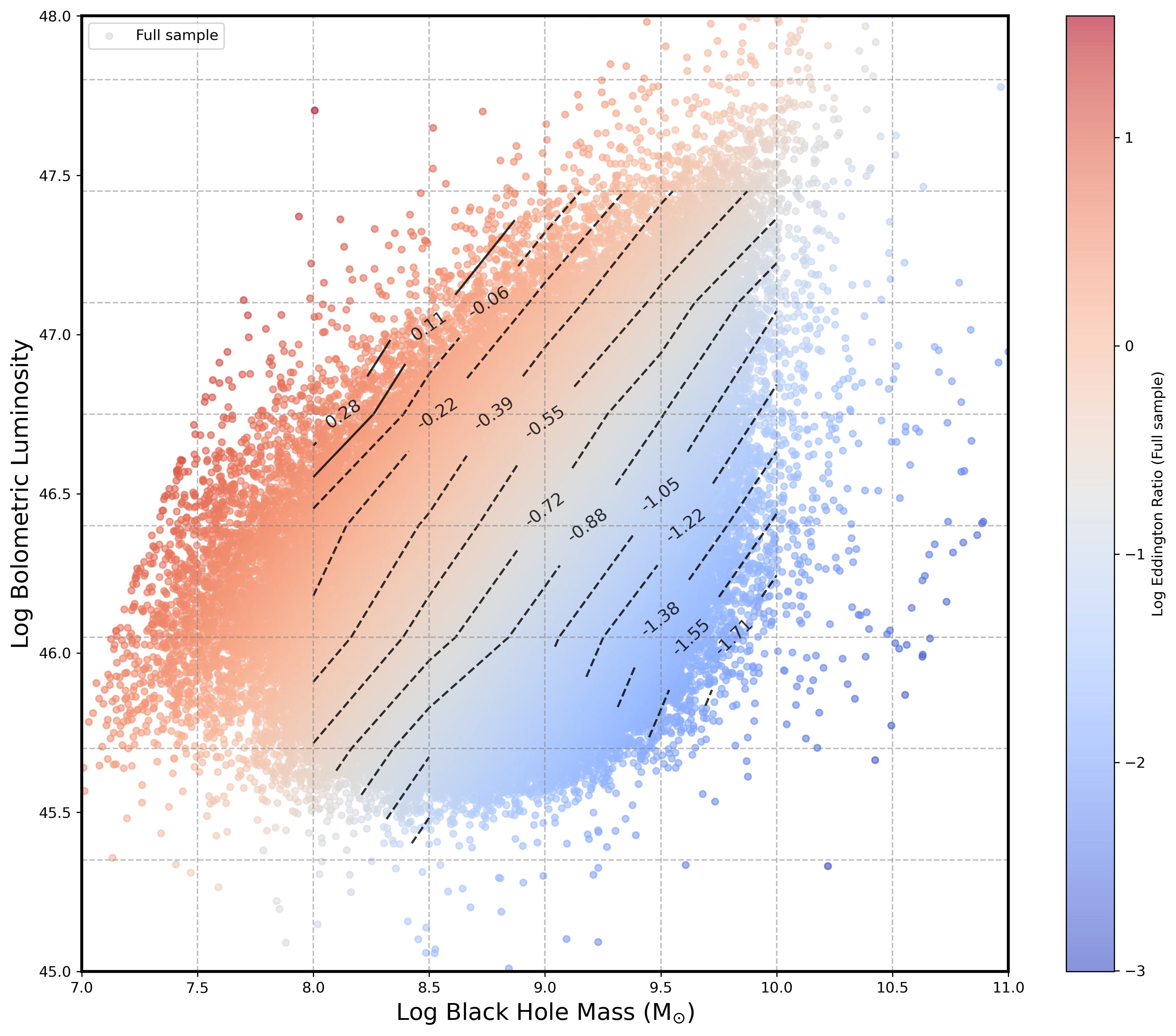} &
    \includegraphics[width=0.5\linewidth]{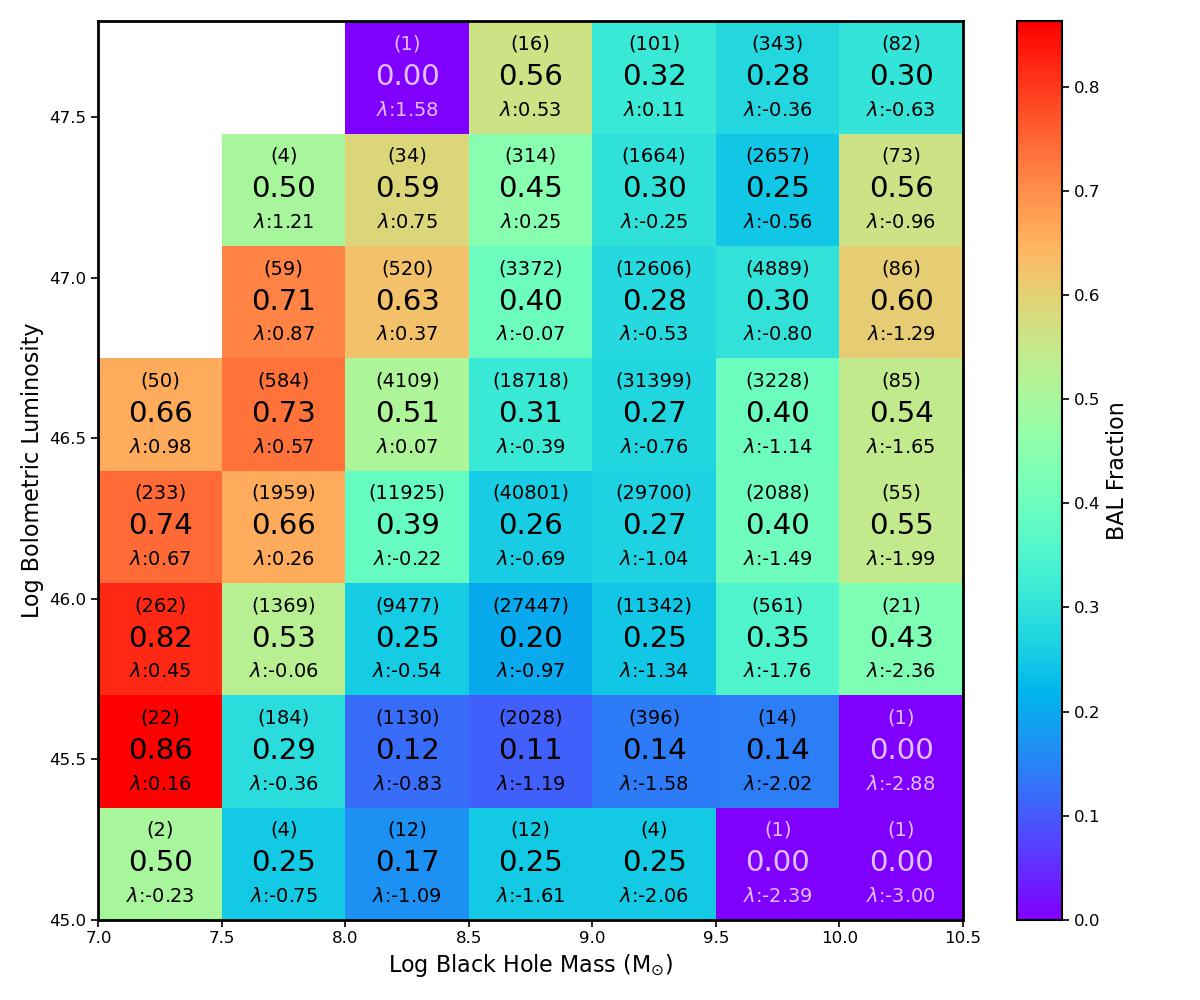}
\end{tabular}
  
    \caption{Left Panel: Scatter plot of black hole mass (log $M_{BH}$) vs. bolometric luminosity (log $L_{bol}$) for the full quasar sample in this study. Data points are color-coded by their log Eddington ratio ($\lambda$).  The black contours indicate the  Eddington ratio distribution of the main sample, with the  Eddington ratio values marked at each contour level.}  Right Panel: Binned grid of log $M_{BH}$ and log $L_{bol}$, showing the BAL fraction in each bin (marked at the center and color-coded). The total number of quasars and the median Eddington ratio in each bin are displayed at the top and bottom of each bin, respectively. 
    \label{fig:bal_frac_MBH_LBOL}
\end{figure*}

The left panel of Fig.~\ref{fig:bal_frac_MBH_LBOL} presents a scatter plot of black hole mass (\(\log M_{\text{BH}}\)) versus bolometric luminosity (\(\log L_{\text{bol}}\)) for a sample of quasars, with data points color-coded by their log Eddington ratio (\(\lambda\)). To enhance statistical significance, we include the full quasar sample with S/N $>$ 3, covering the redshift range 1.5$<$z$<$4.5, as the restricted redshift sample contains fewer sources.   It is evident that low Eddington ratio sources tend to have larger black hole masses as compared to high Eddington ratio sources. To examine trends in BAL fraction with \(L_{\text{bol}}\) and \(M_{\text{BH}}\), we grouped sources into bins of \(M_{\text{BH}}\) and \(L_{\text{bol}}\) and computed the BAL fraction within each bin. The right panel of Fig.~\ref{fig:bal_frac_MBH_LBOL} shows these binned BAL fractions, with color coding indicating the BAL fraction in each cell. Each bin's center shows the BAL fraction, while the top and bottom of each cell display the total quasar count and median Eddington ratio, respectively. Bins with more than 50 sources are marked in black to facilitate better distinction, while bins with fewer than 50 sources are marked in white.

At constant luminosity, moving from higher to lower black hole mass bins increases the Eddington ratio. Along each row (constant bolometric luminosity), the BAL fraction tends to increase for both low and high Eddington ratio bins, with the minimum BAL fraction observed around Eddington ratios of approximately -0.5 to -1 and increasing on either side. Similarly, along each column (constant black hole mass), decreasing luminosity decreases the Eddington ratio, showing a similar trend where the BAL fraction increases on either side of the -0.5 to -1 Eddington ratio range. This trend of increasing BAL fraction is particularly notable in sources with high black hole masses, as Eddington ratios in these bins extend beyond -1.0. In contrast, bins for lower black hole masses lack sufficient sources with Eddington ratios below -1. In conclusion, our results clearly show that the BAL fraction increases for both low and high Eddington ratios, even when controlling for either bolometric luminosity or black hole mass.
\section{Continuum normalization \& BAL identification}\label{app_3}
To measure the absorption parameters, we first normalized each spectrum in our low and high Eddington ratio samples using an  Principal Component Analysis (PCA) reconstruction.  An iterative masking function is applied to exclude regions heavily affected by absorption features. We first performed a Discrete Wavelet Transform using a Daubechies-4 (db4) wavelet mask \citep{daubechies1992}. In the resulting coefficients, we retained only the first two low-frequency components and then reconstructed the continuum. We then masked the regions where the flux falls below the continuum by one standard deviation.  The Daubechies wavelet is particularly effective for capturing sharp discontinuities in data, making it well-suited for isolating broad absorption features that deviate significantly from the continuum. The PCA eigen-spectra were derived from a sample of 55,425 non-BAL quasars in a similar redshift range, with the first five components capturing key spectral features. 

However, visual inspection of the continuum revealed that in cases where absorption lines significantly overlap with emission lines, the PCA continuum can become unphysical. To address this, we adopted an alternative normalization procedure, fitting the continuum using a quasar composite spectrum from \citet{Vandenberk2001}.  The code  fits the masked spectrum to a composite non-BAL quasar model using three adjustable parameters: amplitude, spectral index, and smoothing kernel width. This fitting function scales the composite spectrum in flux, adjusts the spectral slope to match the spectral index of the target quasar, and applies slight smoothing to account for variations in emission line widths between the composite and individual spectra. 

We ultimately performed a visual inspection of both normalized spectra for each quasar and selected the best continuum based on visual assessment. For the majority of sources, both approaches provided similar continuum fits, in which case we selected the PCA-derived continuum. However, in cases where the PCA approach failed to produce a reliable fit, we adopted the composite continuum instead. The visual inspection also facilitated the identification of HiBALs, LoBALs, and FeLoBALs within each sample. In the low Eddington ratio sample, we identified 98 HiBALs, 14 LoBALs, and 2 FeLoBALs, while 7 sources did not exhibit any detectable absorption feature with our continuum normalization methods. Similarly, in the high Eddington ratio sample, we identified 79 HiBALs, 33 LoBALs, and 1 FeLoBAL, with 8 sources showing no absorption feature.   A Fisher’s exact test yielded an odds ratio of 0.34 with a p-value of 0.0029, indicating that LoBALs are  less prevalent in low Eddington ratio sources as compared to high Eddington ratio sources.

To identify BAL regions, we first smoothed the spectra using a 5-pixel window to reduce noise while preserving the absorption features. We then scanned for regions where the flux dips continuously below 0.9, marking these as absorption regions. The start and end wavelengths of each identified absorption feature are recorded, and close absorption components (within 5 pixels of each other) are merged to consolidate overlapping or closely spaced features into a single, continuous absorption region. 

\section{Composite Spectra}\label{subsec:Composite}\label{app_4}
\begin{figure*}
    \centering
    \includegraphics[width=1\linewidth]{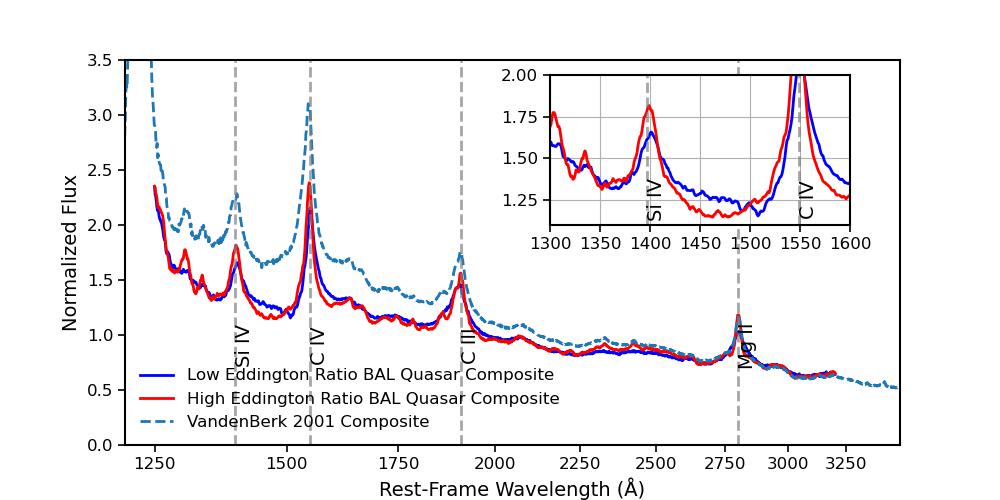}
    \caption{The figure shows the composite spectra for low and high Eddington ratio BAL quasars, smoothed using a Savitzky-Golay filter with a 5-pixel window and third-order polynomial to enhance spectral clarity. The VandenBerk 2001 quasar composite is also shown for comparison. The inset shows a zoomed region of CIV BAL absorption, highlighting that low Eddington ratio sources have BAL features at lower velocities, while high Eddington ratio sources show absorption at higher velocities. }
    \label{fig:composite}
\end{figure*}

We further constructed composite spectra for the 121 low Eddington ratio BAL quasars and 121 high Eddington ratio BAL quasars in our sample. To do this, we resampled all spectra to a common wavelength grid and normalized each spectrum by its mean flux before stacking them to create the composite. Additionally, we tested max-normalization instead of mean normalization and found the results to be unchanged. Fig.~\ref{fig:composite} presents the composite spectra for both low and high Eddington ratio BAL quasars, illustrating the averaged spectral features across the two Eddington ratio regimes in our sample.  We smoothed the flux using a Savitzky-Golay filter with a 5-pixel window and a third-order polynomial to enhance the clarity of spectral features. The resulting composite spectra reveal striking differences between low and high Eddington ratio sources.  Notably, the iron emission in high Eddington ratio sources is significantly enhanced as compared to that in low Eddington ratio sources, aligning with expectations for high Eddington ratio sources in the Eigenvector 1 context. The inset panel displays a zoomed-in view of the \civ\ BAL absorption region, where it becomes evident that low Eddington ratio sources exhibit BAL features at lower velocities, while high Eddington ratio sources show BAL absorption at higher velocities.  This is expected, if the dynamics of outflowing winds in quasars are closely linked to Eddington ratios \citep[e.g.][]{Ganguly2007}. We conclude that in high Eddington ratio sources, the increased radiation pressure more effectively overcomes gravitational forces, accelerating the outflowing material to higher velocities. Conversely, lower Eddington ratios, with relatively weaker radiation pressure, are associated with slower wind velocities.

\end{document}